%% file: main.tex
\documentclass[%
amsmath,amssymb,
aps,prr,
showpacs,
superscriptaddress,
twocolumn
]{revtex4-2}

\usepackage{cancel}
\usepackage{tikz}
\usepackage{hyperref}
\usepackage{graphicx}
\usepackage{xcolor}
\usepackage{dcolumn}
\usepackage{bm}%

\newcommand{\infdiv}{D\infdivx}
\usepackage{mathtools}
\DeclarePairedDelimiterX{\infdivx}[2]{(}{)}{%
	#1\;\delimsize\|\;#2%
}

\graphicspath{{img/}}

\begin{document}
	
	\preprint{APS}
	
	\title{A simple reconstruction method to infer nonreciprocal interactions and local driving in complex systems}
	
	\author{Tim Hempel}
	\affiliation{ 
		Department of Mathematics and Computer Science, Freie Universit\"at Berlin, Berlin, Germany
	}
	\affiliation{ 
		Department of Physics, Freie Universit\"at Berlin, Berlin, Germany
	}
	\author{Sarah A.~M. Loos}
	\email{sl2127@cam.ac.uk}
	\affiliation{ 
		DAMTP, Centre for Mathematical Sciences, University of Cambridge, Wilberforce Road, Cambridge CB3 0WA, UK
	}
	\footnotetext{\hspace{-6mm}* Corresponding author: sl2127@cam.ac.uk}
	
	\date{\today}
	
	\begin{abstract}
		Data-based inference of directed interactions in complex dynamical systems is a problem common to many disciplines of science. In this work, we study networks of spatially separate dynamical entities, which could represent physical systems that interact with each other by reciprocal or nonreciprocal, instantaneous or time-delayed interactions. We present a simple approach that combines Markov state models with directed information-theoretical measures for causal inference that can accurately infer the underlying interactions from noisy time series of the dynamical system states alone. Remarkably, this is possible despite the built-in simplification of a Markov assumption and the choice of a very coarse discretization at the level of probability estimation. Our test systems are an Ising chain with nonreciprocal coupling imposed by local driving of a single spin, and a system of delay-coupled linear stochastic processes. Stepping away from physical systems, the approach infers cause-effect relationships, or more generally, the direction of mutual or one-way influence. The presented method is agnostic to the number of interacting entities and details of the dynamics, so that it is widely applicable to problems in various fields.
		%
	\end{abstract}
	\maketitle
	%
	

	Many effective interactions in natural and synthetic systems have a directed nature, a phenomenon described by a
	variety of scientific disciplines. 
	An example in physics are \textit{nonreciprocal} interactions, defined by the violation of Newton's principle, which ubiquitously occur in nonequilibrium, many-body systems, such as complex plasmas \cite{fortov2005complex,morfill2009complex}, driven soft matter \cite{hayashi2006law}, active matter \cite{ballerini2008interaction,fruchart2021non,loos2023long}, or open quantum systems \cite{metelmann2015nonreciprocal}.
	The particularly distinct edge case of a unidirected or one-way coupling ($A \to B$, $A \not \leftarrow B$) frequently occur in neural networks \cite{hodgkin1952measurement,dayan2005theoretical}. In information theory, a concept of a directed influence are ``\textit{cause--effect}'' relationships between events~\cite{granger1988some}, which is applicable to physical and non-physical systems and widely used in socio-economics and micro and macro-biology. In control theory, the relationship between any \textit{controller} and controlled system, or between any \textit{sensor} and sensed variable, are inherently directed \cite{lozano2022information}.

	From a physics perspective, directionality can be imposed by local driving, e.g., an external driving of only $A$ may result in an asymmetric influence $A\to B$.
	In turn, having directed interactions, is an unambiguous indication for the presence of some sort of driving on the involved physical systems, as nonreciprocity is strictly forbidden in thermal equilibrium \cite{loos2020irreversibility}. 
	Although the concepts of causality, nonreciprocity, and control, stem from different scientific fields, they are inherently interrelated. In this work, we will explore and exploit their underlying connection.

	Directionality of interactions between different local entities has a great impact on their dynamics, thermodynamical properties \cite{ito2013information,loos2020irreversibility}, collective behavior \cite{sanchez2002nonequilibrium,hanai2022non,loos2023long}, and thus functionality. Unraveling directed relationships can therefore be key to understanding the underlying physics and to build suitable models from observed data. A particularly interesting example is neurophysiology, where neural spike train recordings are used to infer functional relationships in human brains~\cite{quinn_estimating_2011}. 
	
	

\begin{figure}
		\centering
		\includegraphics[width=\columnwidth]{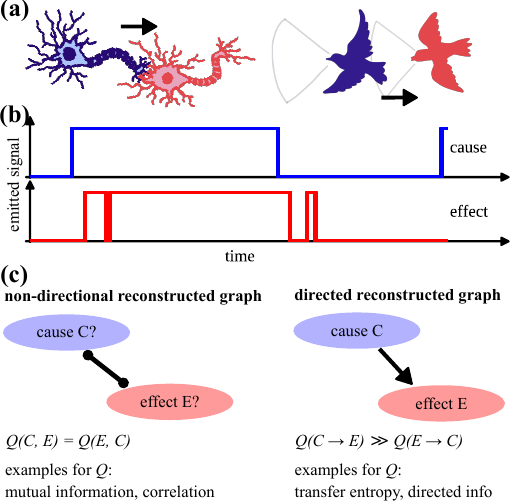}
		\caption{A schematic representation of the problem under consideration.
			\textbf{(a)} 
			Two examples for directed interactions. Left: the unidirectional coupling between to neurons \cite{hodgkin1952measurement,dayan2005theoretical}, Right: the effective nonreciprocal interactions between two birds of a flock, mediated by their visual perception with limited vision cones \cite{ballerini2008interaction}.
			\textbf{(b)} A time-series of a binary delayed channel that serves as example of a directed influence. 
			Note that by definition ``cause'' always precedes ``effect''. Such a time-series could be generated by two entities with unidirectional coupling.
			\textbf{(c)} General types of interaction graphs that could be inferred from the data. 
			A non-directional analysis yields symmetric graph edges and thus no assignment into cause or effect (left). 
			Directed edge analysis enables this assignment (right). 
			The directed influence and the directionality of time, are only captured here.
			$Q$ is a placeholder for the quantity used to infer the graph structure from data, which could be the mutual information for non-directional analysis, or the transfer entropy or the directed information for directed edge analysis.}
		\label{fig:network_type_overview}
	\end{figure}

	Generally, models of directed relationships are \textit{directed graphs}. 
	A directed graph $G=(N,E)$ with nodes $N$ has edges $E$ with the notion of {initial} and {end} nodes \cite{diestel_graph_2017}, so that edges are generally non-symmetric, $e(a, b) \ne e(b, a)$, 
	see Fig.~\ref{fig:network_type_overview} for an illustration. We will interpret this abstract definition as follows: Nodes are local entities that emit a signal which we measure; we attempt to estimate the existence and directionality of the underlying interactions (= edges) from these signals. Edges, which can represent physical or non-physical effective interactions, exchange information between these entities and can therefore be regarded as channels in the information-theoretical sense \cite{shannon1948mathematical}. We generally assume that the local nodes are defined \textit{a priori}.
	We will use the term ``{directed interaction graph}'' to describe the models that we construct from time series data, which are capable of representing directed relationships, such as nonreciprocal interactions or causal influence.

	Two good candidates to infer directed interactions are transfer entropy (TE) \cite{schreiber_measuring_2000} and directed information (DI) \cite{marko_bidirectional_1973,massey_causality_1990}. Both information-theoretical measures quantity directed flow of information and are well-established to determine a causality structure from time-series data. Furthermore, in comparison to Granger causality~\cite{granger_investigating_1969}, they do not build on linear models. Both quantities rely on (estimates of) conditional probability distributions of the observed states. Therefore, they are often used in combination with probability estimators such as continuous tree weighting \cite{jiao_universal_2013} or kernel methods \cite{schreiber_measuring_2000}. The main disadvantage of such estimators is that they are computationally expensive and rely on large data sets.
	
	The key idea of our work is to instead estimate the required probabilities using simple \textit{Markov state models} (MSMs), a powerful framework that has demonstrated its effectiveness in computational biology and statistical physics, see Refs.~\cite{prinz_markov_2011, husic_markov_2018} and references therein.
	MSMs are well-suited to estimate probability distributions in systems with multiple timescales and various states, and can model transition paths between them. They help in reducing the dimensionality by grouping similar states together (i.e., coarsegrain the dynamics into metastable states), thereby making the approach computationally more tractable.
	Moreover, several software packages are available for building and analyzing MSMs.
	For these reasons, MSMs are commonly used to deal with vast high-dimensional, off-equilibrium molecular dynamics data and have successfully been applied to systems as complex as biomolecules \cite{noe_constructing_2009,plattner_complete_2017,paul_proteinpeptide_2017}. 
	
	By combining DI and TE with MSMs, our aim is to present a computationally light yet potent approach for the inference of directed interactions, in possibly very complex, nonequilibrium, and hard-to-sample systems. Compared to other probability estimators, MSMs are based on the assumption of memorylessness, which is one of the main reasons why they can offer a simpler estimation of probabilities than the previously mentioned methods. Furthermore, the presented method uses local, site-specific MSMs, rather than global descriptions of the dynamics, making the approach applicable to high-dimensional systems with sparse sampling~\cite{hempel_independent_2021b,hempel_markov_2022a,olsson_dynamic_2019}.
	
	Our results demonstrate that despite the built-in Markov assumption, our method is capable of inferring the interaction structure. Driving agents can reliably be identified, and even in the presence of explicit memory terms (in the form of time-delayed coupling) the directed interaction graph can be reconstructed.
	
	In the following Methods section, we first introduce MSMs and how they relate to the (conditional and joint) probability distributions needed to compute DI and TE.
	We then give a brief overview of the TE and DI estimators, explicitly show how they can be calculated on the basis of MSMs, and define how we infer the directed interaction graph structure based on them.
	The Results section covers our two example applications, a driven Ising model and an autoregressive model of a driven network of multiple-delayed coupled stochastic processes, which we use to validate the approach.
	Finally, we draw some general conclusions and discuss our results in a broader context.

	\section{Methods}
	\subsection{Probability estimates by MSMs}
	
	The \textit{Markov property} of a time-discrete process $X = \{x_{t_0}, x_{t_1}, x_{t_2}, \dots\}$ with states $x \in S$ at times $t_i$ is defined through the conditional probabilities 
	\begin{equation}
		\label{eq:markov}
		p(x_{t+\tau} | x_t, \linebreak[1] x_{t-\tau}, \linebreak[1] x_{t-2\tau}, \dots) = p(x_{t+\tau} | x_t).
	\end{equation}
	This means that the probability of transitioning from a current state $x_t$ to a future state $x_{t+\tau}$ with $\tau>0$,
	is only determined by $x_t$ and not by the history of the process. The Markov property indicates that the process has no memory of its history.

	A Markov state model (MSM) is based on a given state discretization $S = \{ S_1, \dots, S_n\}$. The key quantity is 
	the transition (probability) matrix $\mathbf{T}$, which gives the ensemble of transition probabilities between all discrete states of a MSM, i.e., 
	\begin{equation}
		\left\{\mathbf{T}(\tau)\right\}_{ij} = p(x_{t+\tau} \in S_j | x_{t} \in S_i).
	\end{equation}
	Importantly, the transition matrix is generally row-stochastic, meaning it has only non-negative entries and columns sum to one: $\sum_j \left\{\mathbf{T}(\tau)\right\}_{ij}  = 1$ (due to conservation of probability). As an approximation to the transfer operator, it propagates probability vectors in time \cite{prinz_markov_2011}, i.e.,
	\begin{equation}
		\bm{p}_{t+\tau}^\top = \bm{p}_{t}^\top  \mathbf{T}(\tau)\, ,
	\end{equation}
	where $\bm p_t$ is a vector describing the probability of each state at time $t$ and $\bm{p}^\top$ denotes its transposed. 
	The first eigenvector with eigenvalue 1 of $\mathbf{T}$ describes the steady-state distribution of the process which does not change under application of $\mathbf{T}$ and encapsulates
	unconditional probabilities \cite{prinz_markov_2011}; it is the stationary distribution of the Markov process $\pi$.
	
	Employing MSMs to a given (sub-)system $X$ (with continuous or discrete configuration space) amounts to approximating the true dynamics at long times by a Markov chain on a discrete space  (given by the chosen discretization $S$). 
	MSMs have become a popular tool for the modeling of molecular dynamics (MD) data, providing a way to combine ensembles of short MD trajectories into a single model.
	MSMs describe ensemble averages as well as kinetic properties. 

	In comparison to the classical MSM approach~\cite{prinz_markov_2011}, we here apply MSMs 
	to arbitrary local entities, i.e., dissecting a global system into smaller, local subsystems \cite{hempel_markov_2022a}. 
	Strictly speaking, this approach is only justified for statistically \textit{independent} subsystems, i.e., disregarding all couplings, 
	a case in which the transition matrix decays into Kronecker products of subsystem transition matrices \cite{hempel_independent_2021b}. In this case, the built-in Markov assumption is valid for each local subsystem. Here, we assume that the partitioning works even in the presence of (weak) couplings, and use the resulting (conditional) probability distributions as approximations.

	We use the conditional and unconditional probabilities estimated from local MSMs for modeling directed interaction graphs between local entities. Local entities that ``emit'' time-series information might be residues of a protein or spins of an Ising chain. We don't explicitly discuss in this work the problem of finding these entities, which correspond to the nodes of the graph, but assume that they are \textit{a priori} defined.  
	As MSMs are defined on a discrete basis, one needs to find suitable discretizations of the configuration space. To carve out the computationally least expensive method, we will employ discretisations $S$ with an as small as possible number of system states $n$.
	
	Once a discretization is defined, we can use local MSMs to estimate the conditional and unconditional probabilities for local entities.
	Joint probabilities for the states of all pairs of nodes are estimated on the discrete joint state space, i.e., by defining all combinatorial states of a pair of nodes using simple rules. 
	In practice, for example, two nodes that both have  two possible states and that are currently in states $n$ and $m$ could be mapped to a joint combinatorial space using the simple assignment rule $(n, m) \rightarrow n + 2\cdot m$.
	Without assuming independence, a new estimate for the dynamics in combinatorial state space is made based on combinatorial discrete trajectories.
	The new estimate yields the combinatorial transition matrix or conditional probabilities $p(x_{t+\tau}, y_{t+\tau} | x_t, y_t)$ and joint stationary distribution $\pi(x, y)$ of two sub-system processes $X$ and $Y$. 
	This approach implicitly makes the assumption that the marginal and joint evolution of single and pairs of nodes in the discretized state space obey a Markovian evolution. 
	Even though this may not be true, in general, the (joint) probabilities can be used as approximations within our framework.
	The great advantage of this approach lies in its simplicity and calculation efficiency.

	In the following, we use upper indices to indicate which processes a probability refers to, e.g., the stationary vector of process $X$ is written as $\pi^X$, and the corresponding vector of the joint space as $\pi^{X,Y}$.

	\subsection{Information-theoretical quantities}
	A common measure for assessing the influence that two entities have on each other is the mutual information between their signals. The mutual information is however, symmetric, i.e., not sensitive to directionality (see Supplemental Material). We will therefore use two related, but directed measures, transfer entropy and directed information. In the following, we will explain how they can be evaluated based on MSM probability estimates. Throughout, we use the base-2 logarithm. This implies a unit of bits for all informatic quantities, which we suppress to ease the notation.
	
	\subsubsection{Transfer Entropy}
	The transfer entropy (TE) from $X$ to $Y$ measures the amount of  information transferred from one stochastic process to another, say from a signal $X$ to a signal $Y$, capturing statistical dependencies and predictive power between them~\cite{schreiber_measuring_2000}. More precisely, it quantifies the reduction of uncertainty about the present values of $Y$ by knowing the history of $X$, given that the history of $Y$ is known. Thereby, it also quantifies the influence the process $X$ has onto process $Y$ .
	
	It is generally defined as \cite{schreiber_measuring_2000}
	\begin{align}
		\text{TE}(X \rightarrow Y) = H(Y_t|Y_{\{t-d:t-\Delta t\}}) \nonumber \\- H(Y_t|Y_{\{t-d:t-\Delta t\}},X_{\{t-d:t-\Delta t\}}),
	\end{align}
	where $Y_{t-d:t-\Delta t}= (Y_{t-d}, Y_{t-d +\Delta t}, ..., Y_{t-2\Delta t},Y_{t-\Delta t})$ denotes the truncated history of the process with memory length $d$ and time step $\Delta t$, and $H(X|Y)$ denotes the conditional entropy being defined as the expectation value $H(X|Y)=-\mathbb{E}[\log_2(p(X|Y))]$.

	The TE thus generally relies on (estimated) probabilities conditioned on the (truncated) history of the random processes. As we use MSM to estimate probabilities, which amounts to the assumption of Markovianity [Eq.~(\ref{eq:markov})] and stationarity, the possible influence from the past is reduced to the present state (i.e., memory length 1). The corresponding conditional probabilities are encoded in the MSM transition matrix elements $\mathbf{T}(\tau)$.
	The resulting (simplified) estimator for TE reads

	\begin{align}\label{TE}
		\text{TE}(X \rightarrow Y) 
		&= \sum_{i, j, k} \pi^{X,Y}_{i,j} \cdot p(y_{t+\tau} \in S_k | x_{t} \in S_i, y_{t} \in S_j) \nonumber\\
		&\cdot \log_2\left[\frac{p(y_{t+\tau} \in S_k | x_{t} \in S_i, y_{t} \in S_j)}{p(y_{t+\tau} \in S_k | y_{t} \in S_j)}\right].
	\end{align}
	In the Supplemental Material, we give a thorough derivation of this expression, starting from the general definition of the TE. Note that $X$ (and $Y$) in Eq.~\eqref{TE} can be placeholders for joint processes, e.g., $X$ may represent multiple nodes of a graph, defined on combinatorial state spaces.

	The transition probabilities of $Y$ conditioned on $X$, $p(y_{t+\tau} \in S_k | x_{t} \in S_i, y_{t} \in S_j)$, are obtained by marginalization of combinatorial transition matrix elements over $x_{t+\tau} \in S_l$.
	Independent transition probabilities $p(y_{t+\tau} \in S_k | y_{t} \in S_j)$ are extracted from the local MSMs.
	Given those results, Eq.~\eqref{TE} can be evaluated and yields pairwise, direction-dependent TE between sub-systems. 
	Please note that this definition is an ensemble average, i.e., utilizes the probabilities that MSMs estimate from the full data set.
	
	Because TE probes dependency in a direction-dependent fashion, it can distinguish between agents that are effectively driving the rest of the system, and those that are effectively responding, and it detects asymmetry in the interactions between subsystems.
	
	We take 
	\begin{align}\label{rule_TE}
		\text{TE}(X\rightarrow Y) \gg \text{TE}(Y \rightarrow X) 
	\end{align}
	as a statistical evidence for a directional influence from process $X$ to process $Y$. 
	
	In this case, $X$ can be interpreted as a transmitter and $Y$ as a receiver of information via a channel, or $X$ as a ``controller'' of $Y$ (or, equivalently, $Y$ as a ``sensor'' of $X$). One could also say that $X$ has a causal influence on $Y$. In the context of physical interactions between two alike particles or spins, this corresponds to a nonreciprocal coupling, with a stronger coupling from $X$ to $Y$ (as we explicitly demonstrate below). 
	It should be noted that not all types of nonreciprocal interactions admit such a structure. Specifically, the special case of the ``perfectly nonreciprocal'' interaction \cite{hanai2022non}, where both interaction strengths are equal but of opposite sign (i.e., $J_{ij}=-J_{ji}$), has no directionality in the current sense.

	\subsubsection{Directed information}
	A closely related measure to infer directed influence is the directed information (DI), generally defined as \cite{massey_causality_1990}
	\begin{align}\label{def:generalDI}
		\text{DI}(X \rightarrow Y) = \sum_{t'=0}^{t} H(Y_{t'}|Y_{\{0:t'-\Delta t\}}) \nonumber \\- H(Y_{t'}|Y_{\{0:t'-\Delta t\}},X_{\{0:t'\}}),
	\end{align}
	where $X,Y$ are discrete time series with time step $\Delta t$.

	In comparison to TE, DI explicitly incorporates the current transition of the process $X$ (the condition involves the current state of $X$, unlike in the TE), i.e., ``instantaneous information exchange''. Another difference is that the history is usually not truncated to a given memory depth $d$. However the greatest difference is that the DI is usually summed over the entire time series as indicated in Eq.~\eqref{def:generalDI}. As a consequence, it can continuously grow for stationary processes, whereas TE reaches a constant value. For our estimation based on MSMs, this difference, however, disappears as we evaluate both measures on Markov chains with memory depth one. Therefore, the only remaining difference between TE and DI is the fact that only DI takes into account the instantaneous information exchange.

	Again, we combine DI with MSMs to estimate the required conditional and unconditional probabilities, which amounts to an underlying assumption of Markovianity and stationarity. In this case, the estimator for DI simplifies to
	\begin{align}\label{DI}
		&\text{DI}(X \rightarrow Y) \hspace{1.75in} \nonumber\\
		&= \sum_{i, j, k, l} \pi^{X,Y}_{i,j} \cdot p(x_{t+\tau} \in S_l, y_{t+\tau} \in S_k | x_{t} \in S_i, y_{t} \in S_j)\nonumber\\
		&\cdot \log_2\left[ \frac{p(y_{t+\tau} \in S_k |x_{t+\tau} \in S_l, x_{t} \in S_i, y_{t} \in S_j)}{p(y_{t+\tau} \in S_k | y_{t} \in S_j)} \right].
	\end{align} 
	Note that $X$ (and $Y$) in Eq.~\eqref{DI} can be composed of more than one atomic sub-process defined on combinatorial state spaces (i.e., each of them can represent the joint process of multiple nodes of a graph).
	Comparing Eq.~\eqref{DI} with Eq.~\eqref{TE} makes again apparent that DI is similar to TE except that its conditional probabilities take into account the current state of the causal variable, which TE does not. 
	Using Bayes' theorem, the corresponding conditioned transition probability can be computed via the joint transition probabilities, specifically by evaluating 
	\begin{align}
		p(y_{t+\tau} \in S_k |x_{t+\tau} \in S_l, x_{t} \in S_i, y_{t} \in S_j) = \nonumber \\
		\frac{p(x_{t+\tau} \in S_l, y_{t+\tau} \in S_k | x_{t} \in S_i, y_{t} \in S_j)}{\sum_k p(x_{t+\tau} \in S_l, y_{t+\tau} \in S_k | x_{t} \in S_i, y_{t} \in S_j)},
	\end{align}
	for every observed transition in the trajectories.
	These probabilities form the transition matrix of $Y$ conditioned on the a transition in $X$.
	
	We take 
	\begin{align}\label{rule_DI}
		\text{DI}(X\rightarrow Y) \gg \text{DI}(Y \rightarrow X)
	\end{align}
	as a statistical evidence for a directional influence from process $X$ to process $Y$.

	Note that DI can be related to Granger causality, a term usually referred to  as a linear model of statistical causality.
	Specifically, linear Granger causality and DI are 
	equivalent
	for jointly Gaussian processes \cite{quinn_directed_2015,amblard_measuring_2009} and Granger causality graphs can be inferred using DI \cite{amblard_directed_2011}.


	\subsubsection{Causally conditioned directed information}
	
	Both TE and DI give statistical evidence for a directional influence. 
	In a network of emitters, information flow might however be more complex. For example,
	there can be so-called ``proxy'' influence via \textit{indirect links}~\cite{quinn_estimating_2011}. For example, say there is no direct influence $A \to B$, but $A\rightarrow C$ and $C \rightarrow B$. Then, there could emerge a statistical dependency and causal influence from $A$ to $B$ through $C$. Such indirect proxy influence $A \rightarrow B$ corresponds to a flow of information and causual influence that would be detected by TE and DI. To reconstruct the actual directed link structure, it is 
	important to differentiate between such indirect and direct links. 
	
	A means of ruling out indirect links is to use \textit{causal conditioning}, where one conditions the DI estimation (or the TE estimation) between 
	two processes $ \text{DI}(X \rightarrow Y)$ on the past of a third process $W$. Specifically, we employ the so-called causally conditioned directed information (CCDI), here written as $ \text{DI}(X \rightarrow Y 
	\| W)$, a tool that has been established for causal inference in recent literature~\cite{quinn_estimating_2011,kramer1998directed,jiao_universal_2013}.
	Importantly, it was shown that a random process $X$ directly causally 
	influences $Y$ with respect to the set of processes $V$, iff~\cite{quinn_estimating_2011} 
	\begin{equation}
		\label{eq:ccdi}
		\text{DI}(X \rightarrow 
		Y \| W) > 0, \qquad \forall~ W \subseteq V \setminus\{X, Y\}.
	\end{equation}
	For directed interactions detected by the asymmetry of TE or DI [Eqs.~\eqref{rule_TE}, \eqref{rule_DI}], we subsequently use the condition~\eqref{eq:ccdi} to test whether they are direct directed links or just indirect ones.
	
	To compute the CCDI, we make use of the identities $\text{DI}(X \rightarrow Y \| W) = H(Y\|W) - H(Y\|X, W)$ and $ \text{DI}(X 
	\rightarrow Y) = H(Y) - H(Y\|X)$, where $H(X\|Y)= -\mathbb{E}[\log_2(p(X\|Y))]$
	denotes the causally conditional entropy. The latter is given by minus the expectation value of the logarithm of the causual conditional probability distribution $p(X\|Y)= \Pi_{i=0}^N p(X_{i\Delta t},| X_{\{0:(i-1)\Delta t\}},Y_{\{0:i\Delta t\}})$ with $i$ running over all discrete time steps \cite{amblard2012relation}. With these identities, we can express the causally conditioned DI as
	\begin{equation}\label{def:CCDI}
		\text{DI}(X \rightarrow Y \| W) =  \text{DI}((X, W) \rightarrow Y) -  \text{DI}(W \rightarrow Y)\,.
	\end{equation} 
	Thus, the estimators for DI described above in Eq.~\eqref{DI}, also readily yield an estimate for the CCDI. 
	The form given in Eq.~\eqref{def:CCDI} further offers the natural interpretation that CCDI is the DI from the time series of the joint process $(X, W)$ flowing to $Y$ reduced by 
	the DI from $W$ to $Y$. 
	Note that if $W$ is an independent Gaussian-distributed random variable, 
	the CCDI given in Eq.~\eqref{def:CCDI}
	readily reduces to $I(X \rightarrow Y)$.
	
	For completeness, we note that, in principle, a causal conditional TE estimator could also be constructed analogously. However, as the CCDI is already established and will prove sufficient for our purposes, we will not employ a CCTE here.

	\section{Application and validation}
	
	\subsection{Ising model}
	Can we correctly identify unidirected coupling, or a single ``driving'' agent from data alone?
	In this section, we corroborate that TE and DI with MSM probabilities can indeed be used to infer directed interaction graphs from time-series data using a nonequilibrium version of the Ising model \cite{ising_beitrag_1925} with nonreciprocal interactions, or local driving.
	We show that indeed, using only data and no additional knowledge about the generating structure, it is possible to retain the structure of the local driving from the data.
	In particular, both TE and DI estimators robustly recover which spin is driven; or, equivalently, the underlying nonreciprocal couplings between the spins. 
	
	We study a one-dimensional Ising chain with $N+1$ spins and periodic boundary conditions, which is subject to a local driving that affects only one spin of the chain (and consider the situation that we do not know which one it is). Specifically, we define the Hamiltonian $\mathcal{H}$ as
	\begin{equation}
		\mathcal{H} = - \frac{1}{2} \sum_{i, j} J_{ij} s_i s_j + H(t) s_{0}
	\end{equation}
	with spins $s_i$, a time-dependent local external field $H(t)$, and a coupling tensor $J_{ij}$. In particular, $J_{ij} = J \delta_{i, i\pm 1}$ couples only nearest-neighboring spins.
	The external field $H$ has infinite strength and exclusively acts on one spin (w.l.o.g., we give this spin the number 0, $s_{0}$) with infinite strength, switching its sign in a completely uncorrelated manner, i.e., $H$ (and thus $s_{0}$) is a Markov jump process, or a dichotomous Markov noise. Due to this infinitely strong field, the spin $s_{0}$ is completely unaffected by its neighbors, yet it exerts an influence on them, i.e. the switching events of $s_{0}$ and $s_{1/N}$ obey a cause--effect relationship.%
	
	This simple example demonstrates the close connection between local driving, nonreciprocity, and cause-effect relationships. Instead of a chain in which a single spin is driven by an finitely strong random field, there is a second fully equivalent way of interpreting the system as a nonreciprocal ($J_{0 j} \neq J_{j 0 }$) spin chain \cite{hanai2022non,loos2023long} with unidirected links to $s_{0}$ (without external field) with a local energy \cite{loos2023long,avni2023non,lima2006ising}
	\begin{equation}
		\mathcal{H}_i = - \frac{1}{2} \sum_{j\neq i} J_{ij}s_i s_j ,
	\end{equation}
	with $J_{0i}=J_{N 0}=0$, and $J_{ij}=J\delta_{i, i\pm 1}$, for all other links, i.e., with nonreciprocal coupling of $s_0$. 
	
	In both interpretations, the spin $s_0$ drives the system out of equilibrium, and we will thus refer to it as the ``driving'' spin.

	\begin{figure}[h]
		\centering
		\includegraphics[width=\columnwidth]{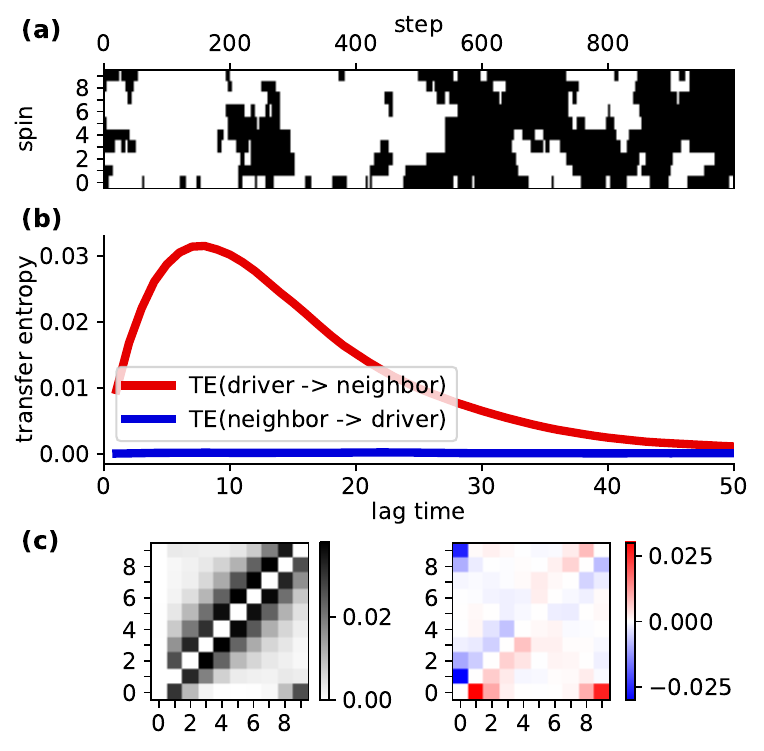}
		\caption{Inference of directed interactions in an Ising chain with 10 spins, one of which is a single driving spin (spin \#0). \textbf{(a)} Excerpt from Glauber dynamics time-series of 1D Ising chain. Here, $\alpha_0=\alpha=0.1$ and $\gamma=0.95$. 
			\textbf{(b)} Transfer entropy (TE) defined in Eqs.~\eqref{TE} estimated between the driving spin and its neighbor, versus the lag time $\tau$. 
			\textbf{(c) Left panel}:  Matrix of pairwise TE, color coded from white (zero bits) to black (maximum TE), for $\tau=10$, which roughly corresponds to the position of the maximum in (b). Note that the 0th column, which denotes TE from any spin to the driving spin, is zero. \textbf{(c) Right panel}: Difference $\text{TE}(X\to Y)-\mathrm{TE}(Y\to X)$. Nonzero values indicate directed links. 
		}
		\label{img:ising_simple}
	\end{figure}
	
	The system is implemented using 
	Glauber-like dynamics with rates~\cite{glauber_timedependent_1963,olsson_dynamic_2019,loos2023long}
	\begin{align}
		w = \frac{1}{2}\alpha_i\left[1-\frac{\gamma_i}{2} s_i (s_{i-1} +s_{i+1})\right],
	\end{align}
	with $\gamma_i =\tanh(2J/kT)$ and $\alpha_i\equiv \alpha$ for all $s_i$ with $i\neq 0$, while we fix for spin $s_0$: $\gamma_0 =0$ and $\alpha_0$.
	The flipping rate of the driven spin $s_{0}$ is thus fully described by the self-transition rate $\alpha_0/2$, and independent of the neighboring spin states. This system has a simple discrete phase space $S$, for which the MSM can be directly employed. The corresponding rate matrix $R$ for single spin flips is approximated by a time-discrete Markov transition probability matrix as $T=\exp(\Delta t\cdot R)$ with a discrete time step $\Delta t$. In order to choose the latter, one aims to find the largest $\Delta t$ for which the spectrum of $T$ and $R$ matches (to speed up calculations). We found that $\Delta t = 1$ is sufficient for this simple system.

	An example of a time-series simulated as a Markov jump process on the transition matrix $T$ is shown in Fig.~\ref{img:ising_simple}a, where the independent driving spin is located at the lowest row ($i=0$). Although the kymograph is noisy, visual inspection gives a hint where the local driving is, as only the flipping of the driving spin is independent, and often followed by an alignment of its environment.
	
	To compute TE, we estimate the needed transition probabilities 
	by employing MSMs with different lag times $\tau$. In Fig.~\ref{img:ising_simple}b) we show the resulting TE between the driving spin and any of its two neighbors.
	Clearly, the directionality of the interaction is correctly identified from \eqref{rule_TE}, even in cases with comparably long lag times.
	Most importantly, we observe no artifacts (e.g., switched directionality) for long lag times.
	
	\begin{figure}[htb]
		\centering
		\includegraphics[width=0.99\columnwidth]{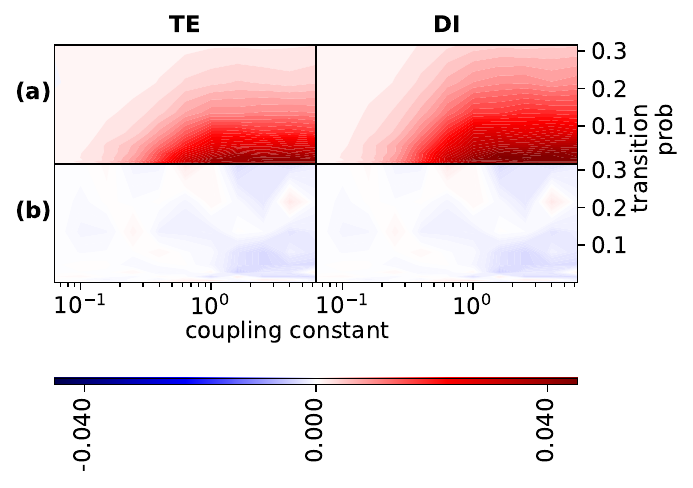}
		\caption{Directionality in Ising model as derived with transfer entropy (TE, 1st column) and directed information (DI, 2nd column).
			For the sake of simplicity, only the differences are presented: $\text{TE}(i \rightarrow {i+1}) - \text{TE}({i+1} \rightarrow i)$ and $\text{DI}(i \rightarrow {i+1}) - \text{DI}({i+1} \rightarrow i)$ evaluated at the time lag where they have their maximal values (encoded as colors).
			Transition probability $\alpha_0$ of the driving spin ($s_0$) on the $y$-axis, coupling constant $J$ on the $x$-axis.
			Row \textbf{(a)}: From driving spin to one of its neighbors. Row \textbf{(b)}: Between two neighboring randomly chosen spins, as a negative control.}
		\label{img:ising_te_di}
	\end{figure}
	
	Furthermore, computing all pairwise transfer entropies in the system (Fig.~\ref{img:ising_simple}c) reveals that no TE goes into the driving 
	spin, and that 
	the two directed links are inferred correctly, while
	no false directionality between any driver-less set of spins is identified. In this case, the directed interaction graph is correctly reconstructed; even without using causal conditioning. 
	The DI brings qualitatively equivalent results as the TE (not shown here).
	
	We conclude that the presented methods thus enables us to correctly detect the local driving (i.e., the unidirected links) purely based on the time-series data with stationary and transition probabilities from MSMs. 
	
	This result however depends on the system parameters. 
	Figure~\ref{img:ising_te_di} demonstrates how TE and DI estimators change with variations on noise level of the driving spin, i.e., the flipping probability (shown on $y$-axis, tuned by changing its rate $\alpha_0$) and the spin-spin coupling strength $J$ ($x$-axis) and driver flipping probabilities $\alpha$. The other spins have fixed $\alpha$'s of 0.1, corresponding to transition probabilities of ~0.005, for sake of comparison.
	We find that the strongest signal for directionality is measured in the region where the noise level of the driving spin is equal as or comparable to the noise on other spins and when the coupling $J$ is strong. This observation can easily be rationalized. First, the signal for directionality inferred from TE and DI decays in the limit where all spins decouple, as expected. Moreover, if the driving spin becomes too noisy such that its flips are too rapid, the system cannot follow (especially if the overall coupling is weak). Concretely, the system fails to respond if the flips are fast as compared to the collective timescales (which one may think of as the ``inertia'' of the spin chain).
	Thus, for extremely low coupling among the spins or very high flipping rate of $s_0$,  our estimators return zero and the local driving cannot be inferred from the data in those regimes by our approach.
	
	In order to further validate our approach, we have compared our results 
	to previous approaches (see Supplemental Material). 
	To this end, 
	we tested the continuous tree reweighing (CTW) probability estimator introduced in Ref.~\cite{jiao_universal_2013}. While the CTW-based estimator is more complex and computationally much more costly, we find that both methods yield comparable results throughout (cf. Supplemental Material, Fig. S1). Furthermore, we tested estimates from reversible MSM~\cite{trendelkamp-schroer_estimation_2015}, which assume time-reversibility of the underlying processes, and find that they degrade our results significantly. This is expected, as time-reversal symmetry is explicitly broken in our nonequilibrium model. In fact, the unjustified reversibility assumption in some cases even leads to false directionality -- something we never observed for the non-reversible MSM estimates presented here (cf. Supplemental Material, Fig.~S1).

	To summarize, TE as well as DI estimators perform well with MSM probabilities, and a driving spin in a 1D Ising chain can reliably be identified purely from the time-series data.

	\begin{figure*}[hbt]
		\centering
		\includegraphics[width=1.1\textwidth]{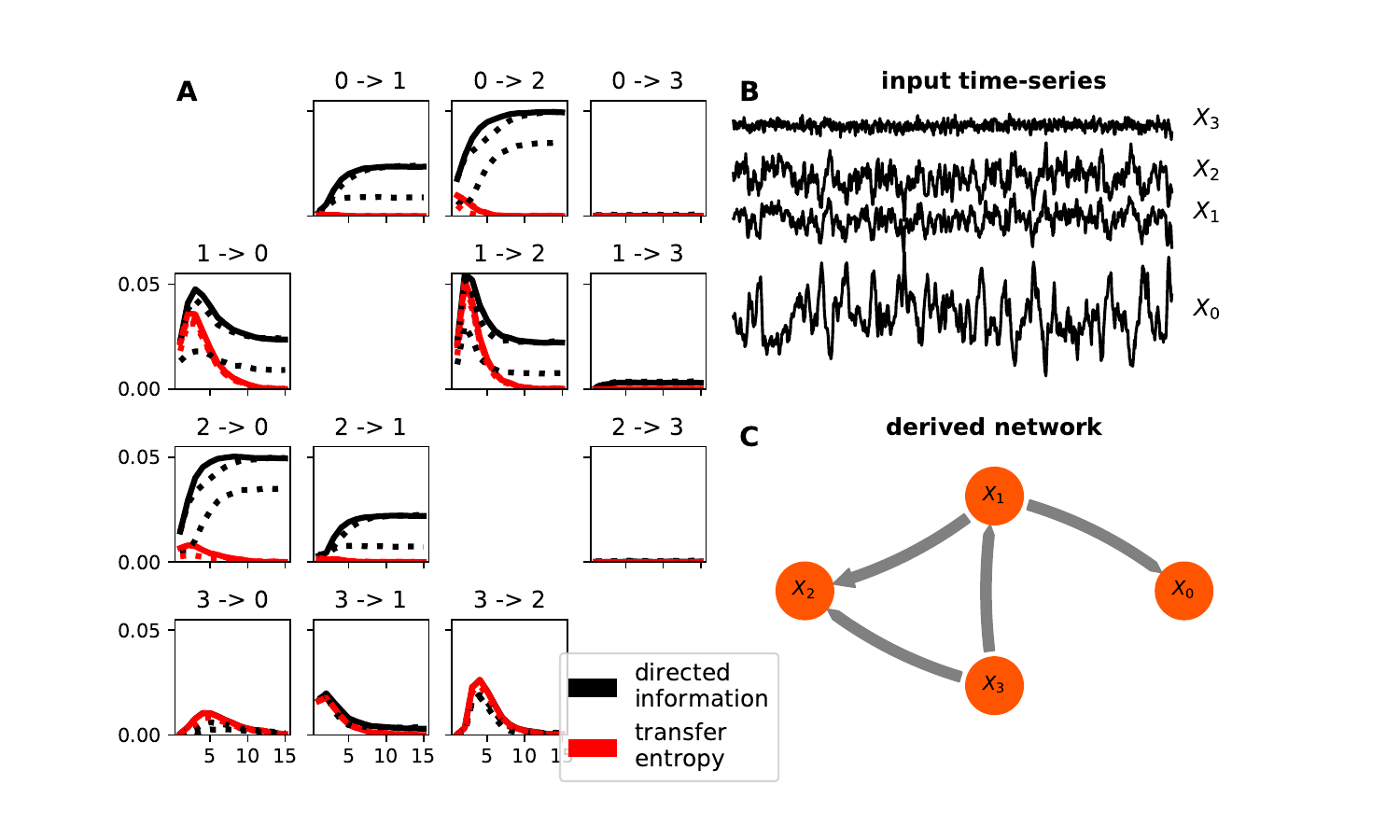}
		\caption{Directed interaction graph reconstruction for a system of delay-coupled linear stochastic processes. \textbf{(a)} Lag time dependence of TE (red) and directed 
			information (black) between processes. 
			Direct, causal links are identified by causally conditioning DI (dotted 
			lines). \textbf{(b)} Excerpt from continuous time-series data of all four processes. \textbf{(c)} Directed interaction graph estimated at a common lag time of 3 steps.}
		\label{fig:ohrnstein_uhlenbeck}
	\end{figure*}
	
	\subsection{Delay-coupled linear stochastic processes}
	In complex systems, cause precedes effect by an undefined time, and 
	interactions may in general be subject to \textit{time delays} \cite{granger1988some}.
	Furthermore, network structures are often more complicated than that of a 1D Ising chain.
	We therefore explore whether the applied simplified probability estimator is suitable to infer interaction graphs from time-series data in a more complex system.
	To test this, we set-up a general autoregressive model, which contains both, processes that influence each other with different {time delays}, and 
	asymmetric, i.e., directed couplings. 
	We show how varying the MSM lag time mitigates the first problem while applying causally conditioned directed information tackles the other.

	Concretely, we generate data for a four-dimensional autoregressive model, with four continuous, delay-coupled random variables $X^{(0)},\,X^{(1)},\,X^{(2)},\,X^{(3)}$, whose dynamics are given by the recurrence relations (or difference equations) \cite{runge_escaping_2012}
	\begin{subequations}\label{example-2}
		\begin{align}
			X^{(0)}_t &= 0.7 X^{(0)}_{t-1} - 0.8 X^{(1)}_{t-1} + \eta^{(0)}_t\\
			X^{(1)}_t &= 0.8 X^{(1)}_{t-1} + 0.8 X^{(3)}_{t-1} + \eta^{(1)}_t\\
			X^{(2)}_t &= 0.5 X^{(2)}_{t-1} + 0.5 X^{(1)}_{t-2} + 0.6 X^{(3)}_{t-3} + \eta^{(2)}_t\\
			X^{(3)}_t &= 0.7 X^{(3)}_{t-1} + \eta^{(3)}_t \,,\end{align}
	\end{subequations}
	where $t$ is a dimensionless discrete time variable with time step $1$, and $\eta^{(i)}$ are independent Gaussian white noises with unit variance. The variables $X^{(i)}$ are coupled among each other in a directed manner (e.g., there are unidirected interactions $X^{(1)} \to X^{(0)}$ and $X^{(3)} \to X^{(1,2)}$), and the interactions involve different {time delays} of length $2,3$ (in addition to the minimal ``delay'' of $1$ time step, which also appears in the self-coupling terms). As a consequence, the stochastic process is non-Markovian and not compatible with thermal equilibrium, i.e., driven. 
	If interpreted as the mesoscopic equations of motions of physical entities, both the directionality of and the time-delays in the coupling correspond to specific forms of nonreciprocity, i.e., the forces among these entities would generically violate Newton's third law.

	An examplary time series~\footnote{The time series was generated using the tigramite software package, version 4.0.0-beta \cite{runge_detecting_2018}.} generated by the model defined in Eqs.~\eqref{example-2} is given in Fig.~\ref{fig:ohrnstein_uhlenbeck}B. The time series display different amplitudes. Clearly, the presence of directed and delayed links is not obvious from a visual inspection of these time series. 
	
	For the MSM estimation, we discretise these continuous (in space) processes into a binary ones. Specifically, we coarsegrain the configuration space from $X^{(i)}\in \mathbb{R}$ to two states $S^{(i)}\in\{-,+\}$; and from $\{X^{(i)},X^{(j\not= i)}\} \in \mathbb{R}\times \mathbb{R}$ to $\{S^{(i)},S^{(j)}\}\in\{(-,-),(-,+),(+,-),(+,+)\}$. 
	Using the thereby obtained MSM estimates of the  probability distributions, we analyze the lag time dependence of DI, TE, as well as CCDI. All estimated measures are shown in Fig.~\ref{fig:ohrnstein_uhlenbeck}A, where DI is depicted by solid black lines, TE by solid red lines and CCDI  with conditioning on either one of the other two processes by two dashed lines.
	By comparing with the generating model, we find that directed links are overall correctly identified by the asymmetry conditions on TE and DI given in Eq.~\eqref{rule_TE} and Eq.~\eqref{rule_DI}. 
	
	In general, DI and TE are further lag time-dependent. 
	We observe that besides their asymmetry, also the functional shapes of TE and DI indicate the type of underlying interaction. A particular characteristic is the occurrence of peaks at certain finite lag times. We find that the occurrence of a peak of a finite time lag (neither at zero nor infinity) is indicative of the presence of a direct link in the underlying model, and that the positions of these peaks mirrors the time delay of a underlying interaction (e.g., from $X^{(1)}\to X^{(0)}$). In contrast, the absence of local maxima at a finite time lag suggests that there are no direct interactions between the nodes (e.g., from $X^{(0)}\to X^{(1)}$). 
	Furthermore, for nodes that are only indirectly connected, but still have an asymmetric TE (such as $X^{(3)}$ and $X^{(0)}$), the CCDI is essentially zero, indicating that this link is only an indirect one. 
	
	A somewhat peculiar observation is that, while TE always decays to zero in the limit of large lag times, DI often converges to a non-zero value. 
	We believe that this is an artifact of our estimator. Recall that the main difference between DI and TE is that only DI is sensitive to instantaneous interactions (zero time delay), which are however not present in the present input data.
	The network structure can still be retained from DI as well as from TE estimates (cf. below).
	CCDI follows the functional form of DI, with the latter forming an upper bound.
	
	The original network structure of the coupled processes can be correctly inferred 
	from the time-series data, even if TE and DI are evaluated at single fixed lag time, e.g., at a lag time of 3 steps used to reconstruct the graph shown in Fig.~\ref{fig:ohrnstein_uhlenbeck}C as an example. To reconstruct such graph,
	we identified directed links by the condition~\eqref{rule_TE} for a fixed $\tau$ and excluded indirect links among them by the condition~\eqref{eq:ccdi}.
	Besides an probability estimation based on a coarsegraining into binary time series (and with built-in Markov assumption), this methods reconstructs the ground-truth network structure with all its directed links.
	
	To go to more complex networks structures (e.g., featuring nonlinear couplings or much longer time delays), a one-fits-all approach to the choice of lag time might not suffice, in particular if the underlying time delays are of various orders of magnitude. However, as we have shown in Fig.~\ref{fig:ohrnstein_uhlenbeck}A, the full lag time-dependence of the inference quantities actually contains much more information than we have used to infer the graph in Fig.~\ref{fig:ohrnstein_uhlenbeck}C, so that a more refined approach that utilizes more of this information could readily be constructed for such cases.
	
	\section{Conclusions}
	We have presented a simple and computationally efficient approach to infer directed couplings such as nonreciprocal interactions from time-series data that combines MSMs with the causal inference tools TE and (CC)DI. 
	We have demonstrated the performance of this approach for a spin model with a single driving spin, and a network of delay-coupled linear stochastic processes. In both cases, our approach correctly identified the nonreciprocal interactions and local driving, and distinguished indirect influence from direct interactions. Thus, the underlying interaction graph structure could be successfully reconstructed. Our method requires no prior knowledge of the system, apart from the prior identification of the dynamical agents, which could represent particles or spins in a physical system.

	A central ingredient of our approach is the approximation to treat local (transition) probabilities of individual parts of a larger complex system as stationary and history-independent, neglecting the memory effects resulting from the coupling to other subsystems. 
	Remarkably, such a probability estimator, which at first glance appears to be an oversimplification, proves to be sufficient to correctly and reliably identify directed connections. Thus, on the level of (local) probability estimates, such memory effects are not essential and can be neglected for the inference of directed interactions or causality. Another interesting insight from this study is that lumping the configuration space into only few coarsegrained states is sufficient to infer the directed interactions.
	
	In the context of physical systems, it is noteworthy that our approach is much simpler than previous approaches to infer nonreciprocal interactions, e.g., based on the response of a system to external perturbations~\cite{takahashi1998analyses} or from spectral analysis~\cite{sametov2023method}, listed and summarized in the introduction of Ref.~\cite{sametov2023method}. Specifically, our method requires no model assumption of the physical forces and mechanisms, no perturbation of the system, and only uses time-series data of the dynamical system states, which are usually experimentally accessible. In turn, different from these more refined approaches, our approach only gives information about the existence and directionality of interactions, without further specifying them.
	
	Application to a broader class of systems than those considered here is easily possible. It would be particularly interesting for future studies to test this approach on more complex types of dynamics, e.g., with long-ranged interactions, or long-ranged memory, and investigate when it will break down. In the context of the present model systems, a next step would be to test the approach on spin systems close to a phase transition, or on networks of diverse nonlinear stochastic processes. Conceptually, it is further straightforward to extend the method to hidden MSMs \cite{rabiner_tutorial_1989}, which is computationally more expensive but amenable to protein dynamics. Indeed, we recently demonstrated that a combination of hidden MSM and TE can be useful to model alosteric coupling~\cite{hempel_coupling_2020a}.
	
	Another interesting question concerns the problem of finite sampling, which is a major challenge for many branches of computational physics. In the current work, only well-converged datasets were used, a case in which no estimation bias is to be expected.
	Generally, due to finite sampling, probabilities and energies must be reweighted to the equilibrium (or nonequilibrium steady state) ensemble, which can be done by kinetic modeling with MSMs. 
	
	\section*{Data Availability}
	To enable experts from different fields to use our methods, we provide the described estimators in a python software package at \url{https://github.com/thempel/information}.

	\section*{Acknowledgments}
	TH acknowledges financial support from Deutsche Forschungsgemeinschaft (DFG, German Research Foundation) through the SFB/TRR 186, Project No.~A12. SL acknowledges funding by the DFG through the Walter-Benjamin Stipendium (Project No.~498288081), and through the Marie Skłodowska-Curie Actions (MSCA) Postdoctoral Fellowship, Guarantee funding undertaken by the UKRI (Grant Ref.~EP/X031926/1). 
	We express our gratitude to Frank No\'e (FU Berlin) for his support and initiative during the early stages of this project. Furthermore, we thank Brooke E. Husic (Princeton University), Simon Olsson (Chalmers University), and Gabriel Schamberg (University of Auckland) for helpful discussions.

	\bibliography{library.bib} 
	
	
	\onecolumngrid
	\vspace*{1cm}
	\hrule height .1pt depth .1pt width \textwidth
	\vspace*{1cm}
	{\textbf{Supplemental Material}}
	\input{Supplemental}

\end{document}

%% file: Supplemental.tex
\section{Mutual information from MSM}
 The mutual information $M$ between two signals $X$ and $Y$ can be expressed in terms of the Kullback–Leibler divergence between the probability distributions of the joint time series, $P^{X,Y}$, and the probability distributions of the individual time series, $P^X$ and $P^Y$,
  \begin{equation}
   M(X, Y) = \infdiv{P^{X,Y}}{P^X P^Y}\, .
  \label{eq:mi_general}
  \end{equation}
 The Kullback-Leibler divergence is defined as $\infdiv{P}{Q} = \sum_i p_i \log(p_i / q_i)$, and $p_i$ and $q_i$ are two probabilities for the same state $i$.
The mutual information can be interpreted as a direct measure of the ``incorrectness'' of the assumption that processes $X$ and $Y$ are mutually independent~\cite{schreiber_measuring_2000}. 
 Throughout, we use the base-2 logarithm. This implies a unit of bits for all informatic quantities, which we suppress to ease the notation.

To estimate the mutual information using our MSM-based approach, we approximate the unconditional probabilities appearing in \eqref{eq:mi_general} with MSM stationary distributions. First, we treat $X$ and $Y$ as (approximately) uncoupled Markov processes that we describe individually by MSMs~\cite{hempel_independent_2021b}. This gives the independent probabilities $\pi^X, \; \pi^X$ from the individual MSMs. Second, via a combinatorial MSM, we obtain the unconditional joint probabilities $\pi^{X,Y}$. Then, 
 \begin{equation}
  M(X, Y) = \sum_{i,j} \pi^{X,Y}_{i,j} \; \log_2\left(\frac{\pi^{X,Y}_{i,j}}{\pi^X_i \pi^Y_j}\right)\,.
  \label{eq:mi}
 \end{equation}

\section{Directed information flow quantities from MSM}

\paragraph{Transfer Entropy}
The definition of transfer entropy by Ref.~\cite{schreiber_measuring_2000} is a Kullback entropy. We can therefore re-write it in terms of the combinatorial MSM stationary distribution $\pi$ and marginalized transition matrix, and find
\begin{align}
	\text{TE}(X \rightarrow Y) &= \sum_{\substack{x_{i-1} \in X \\ y_{i-1}, y_i \in Y}} p(y_{i}, y_{i-1}, x_{i-1}) \log_2\left(\frac{p(y_{i} | y_{i-1}, x_{i-1})}{p(y_{i}|y_{i-1})} \right) \\
	&= \sum_{\substack{x_{i-1} \in X \\ y_{i-1}, y_{i} \in Y}} \pi(x_{i-1}, y_{i-1}) p(y_{i}|x_{i-1}, y_{i-1}) 
	\log_2\left(\frac{p(y_{i} | y_{i-1}, x_{i-1})}{p(y_{i}|y_{i-1})} \right)\label{si:te}
\end{align}

\paragraph{Directed Information}
To achieve an equivalent formalism, we have defined directed information with a Kullback entropy as it is the case for transfer entropy.
A connection to the usually used definition of directed information can be drawn upon time-averaging. 
We note that we assume probabilities to be stationary (i.e., do not change with the time index $i$). Under this assumption, we find
\begin{align}
	\text{DI}(X \rightarrow Y) &= \frac{1}{n} \sum_{i=1}^n \infdiv{P(y_i|Y^{i-1}, X^i)}{P(y_i|Y^{i-1})}\\
	&= \frac{1}{n} \cdot n \cdot \infdiv{P(y_i|Y^{i-1}, X^i)}{P(y_i|Y^{i-1})}\\
	&= \sum_{\substack{x_{i-1}, x_i \in X \\ y_{i-1}, y_i \in Y}} p(x_i, x_{i-1}, y_i, y_{i-1})
	 \log_2\left(\frac{p(y_i | x_i, x_{i-1},y_{i-1})}{p(y_i | y_{i-1})}\right)\\
	 	&= \sum_{\substack{x_{i-1}, x_i \in X \\ y_{i-1}, y_i \in Y}}  \pi(x_{i-1}, y_{i-1}) p(x_i, y_i | x_{i-1}, y_{i-1})
	 \log_2\left(\frac{p(y_i | x_i, x_{i-1},y_{i-1})}{p(y_i | y_{i-1})}\right)\label{si:di}
\end{align}
The letter $D$ denotes the Kullback entropy. We note that our definition slightly differs from the original one presented by Ref.~\cite{massey_causality_1990} which uses a Kullback-Leibler divergence instead of the Kullback entropy.

\paragraph{Systems with no instantaneous information exchange }
Please note that in the case of no instantaneous information exchange, directed information and transfer entropy are identical.
In this case, the conditional probability simplifies to $p(y_i | x_i, x_{i-1},y_{i-1}) = p(y_i | x_{i-1},y_{i-1})$ and the sum over $x_i$ can be marginalized over the transition matrix elements $\sum_{x_i}p(x_i, y_i | x_{i-1} y_{i-1}) = p(y_i | x_{i-1}, y_{i-1})$. 
Therefore, directed information becomes
\begin{align}
	\widetilde {DI}(X \rightarrow Y)
	&= \sum_{\substack{x_{i-1} \in X \\ y_{i-1}, y_i \in Y}} \pi(x_{i-1}, y_{i-1}) 
	p(y_i |x_{i-1}, y_{i-1}) \log_2\left(\frac{p(y_i | x_{i-1},y_{i-1})}{p(y_i | y_{i-1})}\right)\\
	&= \widetilde {TE}(X \rightarrow Y).
\end{align}

\section{Comparison with other estimators}

\begin{figure}[h!]
	\centering
	\includegraphics[width=\textwidth]{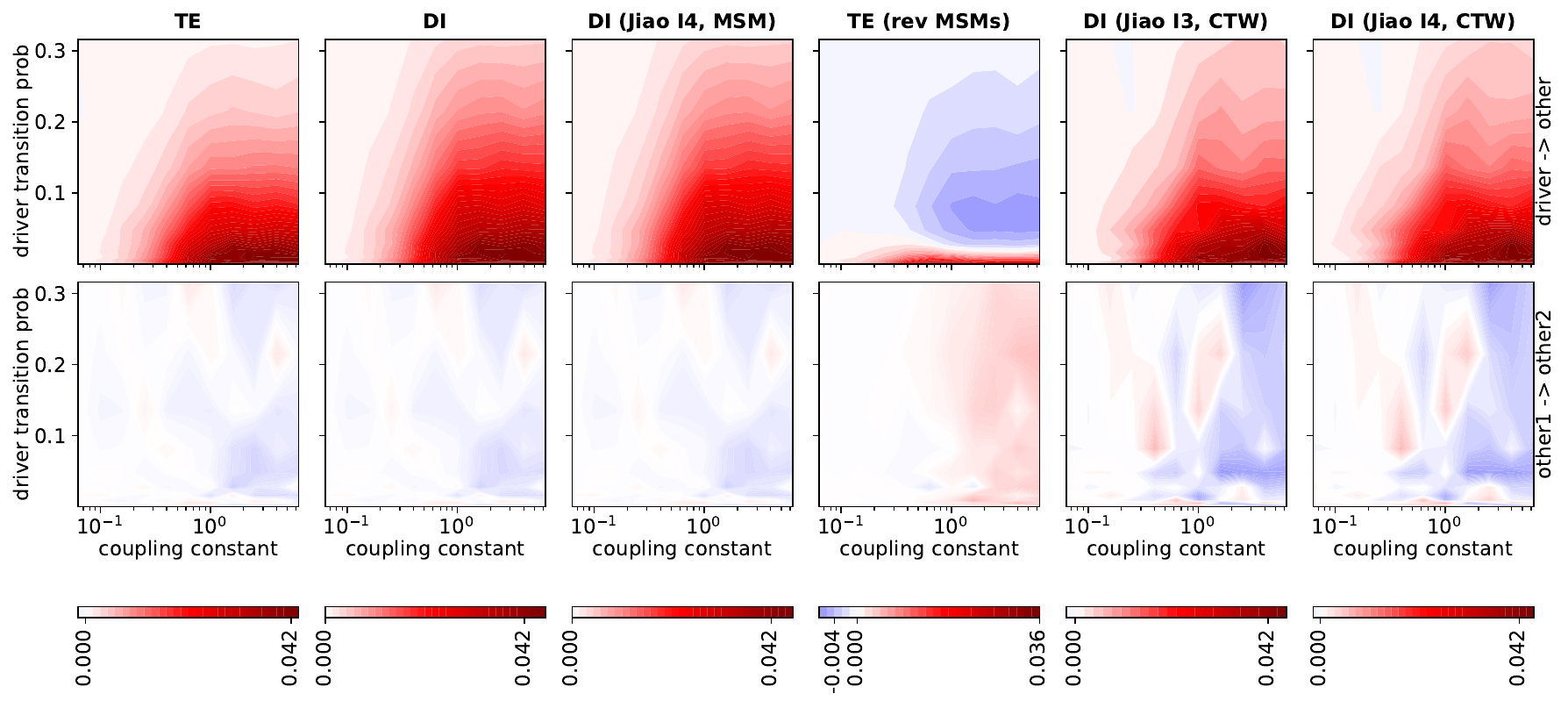}
	\caption{Directed information and transfer entropy in the driven Ising system, computed with different probability estimators. For the sake of readability, only the difference between forward and backward computation is shown. 
\textbf{Top row:} From independent spin to its neighbor.
\textbf{Bottom row:} Between two neighboring random spins as a negative control.
\textbf{First column:} transfer entropy (TE) with MSM probabilities [as defined in Eq. \eqref{si:te}]. 
\textbf{Second column:} directed information (DI) with MSM probabilities [as defined in Eq.~\eqref{si:di}]. 
\textbf{Third column:} DI using the directed information estimator $I_4$ as defined in Ref.~\cite{jiao_universal_2013}, with MSM probabilities.
\textbf{Fourth column:} TE [as defined in Eq.~\eqref{si:te}] with probabilities stemming from a reversible MSM estimator \cite{trendelkamp-schroer_estimation_2015}. 
\textbf{Last two columns:} directed information estimators $I_3$ and $I_4$ as defined in Ref.~\cite{jiao_universal_2013}, with probabilities stemming from the CTW algorithm presented in the same paper.
}
	\label{fig:ising_grid}
\end{figure}

To validate our implementation, we have estimated directed information and transfer entropy with different probability estimators, including the context tree reweighing (CTW) algorithm presented by Ref.~\cite{jiao_universal_2013}.
We used the driven Ising chain as our test system (see Main for definition and parameters of this model). Figure~S\ref{fig:ising_grid} shows the corresponding results for different values of coupling constant and driver transition probability, for the same parameters as in Fig.~3 of the main. 
We find that the directed information and transfer entropy estimated by our approach qualitatively agree with the CTW estimates. In contrast, using a reversible MSM estimator \cite{trendelkamp-schroer_estimation_2015} produces false results. This is expected, as the latter is build on the assumption of time-reversibility, which is explicitly broken in this nonequilibrium system.

